\documentstyle[times]{mn}
\input{psfig}

\def\slHI{{\sl H\,\small\sl I}}

\def\oii{[O{\,\small {II}}]$\lambda3727$}
\def\oiii{[O{\,\small {III}}]$\lambda5007$}

\newcommand{\HI}{H{\,\small I}}
\newcommand{\figHI}{H{\,\scriptsize I}}
\newcommand{\tspin}{$T_{\rm spin}$}

\newcommand{\kms}{km\,s$^{-1}$}

\newcommand{\Lsun}{{$L_\odot$}}
\newcommand{\ltae}{\raisebox{-0.6ex}{$\,\stackrel
{\raisebox{-.2ex}{$\textstyle <$}}{\sim}\,$}}

\begin{document}

\title[H{\,\normalsize\it I} absorption in  radio galaxies]{H{{\,\LARGE\bf I}} 
absorption in radio galaxies: effect of orientation or interstellar
medium?\thanks{Based on observations with the Australia Telescope Compact
Array, with the Very Large Array and with the Westerbork Synthesis Radio
Telescope.}}

\author[Morganti et al.]{R. Morganti$^{1}$\thanks{Email: morganti@astron.nl}, T.A.
Oosterloo$^{1}$, C.N. Tadhunter$^{2}$, G. van Moorsel$^{3}$, 
\and
N. Killeen$^{4}$, K.A. Wills$^{2}$
\\
$^1$ Netherlands Foundation for Research in Astronomy, Postbus 2, 7990
AA Dwingeloo, The Netherlands \\
$^2$ Dept.  Physics, University of Sheffield,
Sheffield S3 7RH, UK \\
$^3$ National Radio Astronomy Observatory, Socorro,
NM 87801, USA \\
$^4$ Australia Telescope National Facility, CSIRO, P.O. 
Box 76, 2121 Epping NSW, Australia}

\date{Accepted~~, Received~~}

\maketitle

\begin{abstract}
  
A search for \HI\ absorption has been made in 23 radio galaxies selected from
a complete sample.  The observations were made with the Australia Telescope
Compact Array (ATCA), with the Very Large Array (VLA) and for one galaxy with
the Westerbork Synthesis Radio Telescope (WSRT).  In five galaxies \HI\
absorption was detected.  We investigate how the detection rate is distributed
among galaxies with different radio and optical properties.  Among the
Fanaroff-Riley (FR) type I radio galaxies, only one object (10$\%$ of total)
was detected in \HI\ absorption.  The \HI\ absorption in these objects is
likely to come from a nuclear disk, as found for other galaxies of this type
(e.g.\ NGC 4261 and Hydra A).  The low detection rate is consistent with the
hypothesis (as suggested by optical and X-ray data) that the ``standard'' pc
scale, geometrically {\sl thick} torus is not required in low-luminosity radio
galaxies.  This is consistent with earlier optical work.  In the case of FR
type-II powerful radio galaxies, no \HI\ absorption has been detected in broad
line radio galaxies, while three out of four narrow-line radio galaxies have
been detected (the one non-detection having quite a high upper limit).  All
these are compact or small radio galaxies.  To first order this is consistent
with the predictions of the unified schemes, assuming that the \HI\ absorption
is due to an obscuring torus.  However, the indications of this being the only
cause of the absorption are not very strong.  In particular, we find that in
two of the three detected objects that the \HI\ is blueshifted compared to the
systemic velocity.  In the third galaxy (PKS~1549--79) two redshift systems
(from the optical lines) are found.  The uncertainty in the systemic velocity
derived from optical lines is discussed.  Finally, by also considering data
available in the literature, we find a tendency for radio galaxies with a
strong component of young stellar population and far-IR emission to show \HI\
absorption.  The overall richer ISM that is likely to be present in these
galaxies may be a factor in producing the absorption. 

\end{abstract}


\section{Introduction}

The near-nuclear regions of powerful radio galaxies are crucial for
understanding the anisotropy of the radiation field and the fuelling of the
activity, but currently very little is known about the distribution and
kinematics of the gas close to the nucleus.  One of the few ways to improve
our knowledge is by studying the \HI\ absorption against their radio nuclei. 
The advantage of this approach is that observations of the \HI\ absorption
against radio continuum sources open the possibility of detecting relatively
small quantities of \HI\ even in high redshift objects. 

\HI\ absorption has been found in a large number of different types of AGN.  
By looking at the properties of the absorption, it appears that this absorption
can be caused by a variety of gaseous components.

For most radio galaxies in which \HI\ absorption has been detected, (see for
example Cygnus A, Conway \& Blanco 1995 and Conway 1998; NGC 4621, Jaffe \&
McNamara 1994 and van Langevelde et al.\ 2000; Hydra A, Dwarakanath et al.\
1995, Taylor 1996), it has been suggested that a disk with radius of the order
of   $\leq
100$ pc can be the cause for such absorption.  In the case of Cygnus~A VLBA
observations (Conway 1998) show that the absorption is occurring against the
counter-jet, and that it is elongated spatially in a direction perpendicular
to the radio jet.  This strongly suggests that the \HI\ absorption is due to a
circumnuclear torus that blocks the view of the quasar nucleus of Cygnus A. In
the case of NGC~4621, the \HI\ absorption on the VLBI scale (van Langevelde et
al.\ 2000) appears against the counter-jet at a projected distance from the
nucleus of only $\sim$5 pc.  The cause of the absorption is believed to be a
{\sl thin} disk ($\sim$1.3 pc  thick).  
Finally, in Hydra~A (Taylor 1996) the
absorption is likely to arise from a disk with thickness of $\sim$30 pc.  All
of these structures could be part of a continuous structure which spans all the
way from a many-kpc scale disk down to the classical accretion disk of a few
tens of parsec in size.  This has also been found to be the case for Seyfert
galaxies (Gallimore et al.\ 1999).

A class of radio sources that appears to be particularly suitable for
detecting and studying the \HI\ absorption are Compact Symmetric Objects (CSO)
and Compact Steep Spectrum (CSS) radio sources, as noted by Conway (1996).
These are strong radio sources, a few kpc (i.e.\ sub-galactic) in size and are
thought to be very young ($\ltae 10^4$ yr).  It is claimed that in these
objects the most likely cause of the \HI\ absorption is also
a nuclear disk/torus.
However, the interpretation of some of these absorption profiles relies on the
value of the systemic velocity that often is very uncertain.

Evidence of very broad absorption ($\sim$400-500 \kms, compared to the 50-200
\kms\ found in general in radio galaxies) has been found in luminous infrared
galaxies (Mirabel 1989) like 4C~12.50 and 3C~433.  In particular, Mirabel
(1989) noted that for 4C 12.50 the shape of the \HI\ absorption profile bares
a striking similarity with that of starburst galaxies like Arp~220 (Mirabel
1982), suggesting a link between radio-IR spirals and powerful radio galaxies
in the early stages of their evolution (like 4C 12.50).

The interaction between the radio plasma and the interstellar medium can also
play a role.  In the Seyfert galaxy IC~5063, very broad ($\sim$700 \kms) \HI\
absorption, blueshifted with respect to the systemic velocity (Morganti et
al.\ 1998, Oosterloo et al.\ 2000) has been observed coincident with a radio
lobe about 1 kpc from the nucleus.  This is clear evidence for a strong
interaction between the radio plasma and the ISM in the galaxy, causing the
cold gas to be 'blown out' from the centre of the galaxy at large velocities.
Evidence for interaction between the radio plasma and the ISM has been found
in at least two radio galaxies: the superluminal object 3C 216 (Pihlstr\"om et
al.\ 1999) and 3C 326 (Conway et al.\ in preparation).
On even larger scale ($>$ 10 kpc), \HI\ absorption against the radio lobes has
been found in the radio galaxy Coma~A (Morganti et al. 2000a). 

In some galaxies (e.g.\ Centaurus~A, van der Hulst et al.\ 1983; NGC~1052 van
Gorkom et al.\ 1986), additional narrow profiles are clearly seen, redshifted
with respect to the systemic velocity by a few hundred \kms.  These profiles
could correspond to gas clouds falling towards the AGN.  Van Gorkom et al.\
estimated that these infalling clouds could provide the right amount of fuel
for the AGN.

It is clear that radio galaxies show a diversity of \HI\ absorption
properties. However, most previous studies have been based on heterogeneous
samples of objects, often including the most spectacular objects in a given
redshift range. In order to establish the \HI\ absorption properties of the
general population of radio galaxies it is important to study a well-defined
sample.

In an attempt to study the presence and properties of the \HI\ absorption in
radio galaxies in a more systematic way and to investigate whether the
presence of the absorption can be related to other characteristics of the
galaxies (e.g.\ orientation, as one would expect if the nuclear disk/tori are
the cause of absorption) we have searched for \HI\ absorption in a sample of
radio galaxies for which a wealth of information (from radio, optical and
X-ray observations) is available.

Here we present the results obtained for 23 radio galaxies observed so far.
In \S 2 we describe the observations that were made either with the Australia
Telescope Compact Array (ATCA) or with the Very Large Array (VLA) depending on
the declination of the objects and for one galaxy with the Westerbork
Synthesis Radio Telescope (WSRT). In \S 3 we describe the results obtained and
we discuss in more detail the galaxies in which \HI\ absorption is found.  In
\S 4 we discuss the occurrence of the \HI\ absorption in the context of the
overall characteristics of the sources, in \S 5 we discuss the difficulties in
the interpretation of the data due to uncertainties in the systemic velocities
derived from optical emission lines.  Finally, in \S 6 we discuss the likely
origin of the absorption in the galaxies of our sample.

In this paper we assume $H_\circ = 50$ \kms Mpc$^{-1}$ and $q_\circ = 0$.

%
%%% Table 1
%
\begin{center}
\begin{table*}
{\sc Table 1.} {Radio \& optical characteristics of the observed galaxies}  
\smallskip   

\begin{tabular}{ccccccccc} \hline \hline
 Name   & Other    &    V$_{\rm hel}$  &  Radio & $\log P_{\rm tot}$ & $\log
P_{\rm core}$ & $R$
& Optical  \\
         & name   &  km s$^{-1}$ & Morph. & 2.3 GHz  & 2.3 GHz  &  2.3 GHz & type$^1$ \\ \hline
0034--01 & 3C 15     & 21885   &  1/2 & 25.80  & 23.77   &  0.010 & WLRG   \\
0035--02 & 3C 17     & 65860  &  2   & 27.00  & 26.00   &  0.113 & BLRG   \\
0055--01 & 3C 29     & 13427   &  1   & 25.50  & 23.70   &  0.016 & WLRG   \\
0123--04 & 3C 40     &  5396   &  1   & 24.67  & 23.05   &  0.025 & WLRG   \\
0305+03  & 3C 78     &  8590   &  1   & 25.30  & 24.33   &  0.121 & WLRG   \\
0325+02  & 3C 88     &  9060   &  2   & 25.11  & 23.71   &  0.042 & WLRG   \\
0428--53 &          & 11400   &  1   & 25.39  & 23.61   &  0.017 & WLRG   \\
0518--45 & Pic~A     & 10510   &  2   & 26.21  & 24.62   &  0.027 & BLRG   \\  
0521--36 &          & 16590   & ...  & 26.23  & 25.25   &  0.117 & BL Lac  \\
0620--52 &          & 15320   &  1   & 25.40  & 24.16   &  0.062 & WLRG   \\
0625--35 &          & 16367   &  1   & 25.59  & 24.86   &  0.227 & WLRG   \\
%0859--25 &          &  0.305  &  2   & 27.26  & $<$24.52 & $<$0.002 & NLRG \\
1246--41 & NGC~4696 &  2958   & ...  & 23.90  & 21.73   &  0.007 & WLRG   \\
1251--12 & 3C~278    &   4497  &  1   & 24.65  & 22.84   &  0.016 & WLRG   \\
1318--43 & NGC~5090 &   3421  &  1   & 24.22  & 23.34   &  0.153 & WLRG  \\   
1333--33 & IC~4296  &   3761  &  1   & 24.87  & 23.15   &  0.019 & WLRG   \\
1549--79 &          & 45672$^2$ & CFS  & 26.59  & 26.34   &  1.310 & NLRG   \\
1637--77 &          &  12801  &  2   & 25.45  & 24.04   &  0.040 & WLRG  \\   
1717--00 & 3C~353    &   9120  &  2   & 26.16  & 23.53   &  0.002 & WLRG   \\
1814--63 &          &  19350$^2$  & CSS  & 26.14  &  ...    &   ...  & NLRG   \\
%1938-15  &         &   0.452 &  2   & 27.68  &  ...    &   ...  & BLRG   \\
1949+02  & 3C~403    &  17688  &  2   & 25.77  & 23.58   &  0.007 & NLRG   \\
2152--69 &          &   8476  &  1/2 & 25.79  & 24.28   &  0.031 & BLRG   \\
2221--02 & 3C~445    &  16848  &  2   & 25.71  & 24.12   &  0.027 & BLRG   \\
2314+03  & 3C~459    &  65924  &  2   & 26.78  & 26.24   &  0.394 & NLRG    \\
\hline
\hline
\end{tabular}

$^1$ NLRG = narrow line radio galaxy, BLRG = broad line radio galaxy, WLRG =
weak line radio galaxy \\
$^2$  new values, see text for details. \\

\end{table*}
\end{center}

\section{The Sample and the Observations}

The observed radio galaxies have been selected from the 2-Jy sample of radio
sources (Wall \& Peacock 1985, Tadhunter et al.\  1993, 1998; Morganti et al.\ 
1993, 1999 and references therein).  This sample includes strong (S$_{2.7\,
\rm{GHz}}>$ 2 Jy) radio sources with declination south of $10^\circ$ and
redshift less than 0.7.  Although the aim was to have a ``complete'' sample of
radio galaxies observed in
\HI, the actual observed sample has been constrained by three technical
limitations: 1) the observed objects have a radio core typically stronger than
$\sim$60 mJy (one exception is the galaxy PKS 1949+02 (3C~403)); 2) the
redshift limit reachable with the available receivers is between 0.15 and 0.22
(depending on the telescope used); 3) even inside this range a few objects
could not be observed because their redshifted \HI\ would fall in spectral
regions with strong known radio interference.  While the first two constraints
have introduced some bias on the selection of the observed objects, the latter
has just randomly affected the sample.

The radio and optical characteristics of the observed galaxies are summarised
in Table 1, mainly taken from Morganti et al. (1997).  In Table 1 we also
included the values of $R$ (defined as the ratio of core to extended radio
fluxes and considered to be a orientation indicator).  We have observed nine
Fanaroff-Riley (FR) type-I galaxies objects, eight FR type-II, two
intermediate morphology galaxies, one BL Lac, and three compact radio galaxies
(but one of them is the very low radio power galaxy NGC~4696).  As derived
from our optical spectra (Tadhunter et al.\ 1993, 1998), four galaxies are
narrow line radio galaxies (NLRG), four are broad line radio galaxies (BLRG)
and the remainder have weak optical emission lines (WLRG, EW$_{\rm [OIII]} <$
10\AA).  Note that some of the FR type-II selected have weak emission lines.
Because of the limitations mentioned earlier, most of the objects in our
sample are FR type-I radio galaxies or broad-line radio galaxies (that are
known to have, on average, stronger radio cores compared to the narrow-line
galaxies, Morganti et al.\ 1997).  Also the large number  of WLRG FR-II galaxies is
due to the fact that they have relatively strong cores (Morganti et al.\ 
1997).

Almost all observations were made either with the ATCA or with the VLA
depending on the declination of the objects.  We have used the 6~km
configuration of the ATCA and the B-array configuration of the VLA.
The spatial resolution of the ATCA and VLA data is typically between 6 and 10
arcsec.  The log of the observations is given in Table 2.  PKS 1333--33 was
observed with both instruments.  The ATCA observations of PKS~2152--69 are
part of a larger study of the radio continuum properties of this object
(Morganti et al.\ in prep). One galaxy (3C 459) was also observed with the
WSRT. The system temperature of the new 21-cm frontends used at the WSRT is  low
($\sim 27$ K) even at the relatively low
frequency of the redshifted \HI\ in this object (1165 MHz), therefore allowing
us to better estimate the characteristics of the absorption in this galaxy.

With the ATCA we were able to observe galaxies up to a redshift $z \sim 0.15$,
corresponding to 1235~MHz.  In the ATCA observations we used a 16 MHz
bandwidth and 512 channels in order to be sure to have both a large velocity
range and good velocity resolution.  The velocity resolution obtained is $\sim
6.6$ km s$^{-1}$ before Hanning smoothing.  Each object was observed a number
of times (for approximately 20 minutes each time) throughout a 12~h observing
period in order to obtain as much as possible uniform {$uv$} coverage.  As
bandpass and flux-density calibrator we have used PKS~1934--638, assumed to
have (according to the latest result by Reynolds, 1996) 15.1 and 14.9 Jy at
1235 and 1405~MHz, the lowest and highest frequencies used.  We spent between
10 min and 1 hour on the bandpass calibrator, depending on the flux of the
core of the observed sources.  The log of the ATCA observations is given in
Table~2a. 

The VLA observations were made with a 6.25 MHz bandwidth and 64 channels.  The
spectral resolution obtained is $\sim$21 \kms.  Most of the observed radio
galaxies have  $z < 0.1$, but we also included two objects with $z \sim 0.22$.
The observing time and frequency for each galaxy are given in Table~2b.

The WSRT observation of 3C~459 was made using the DZB backend, 10 MHz
bandwidth 256 channels given about 12 \kms\ velocity resolution. This galaxy
was observed for 3.6 h.

The spectral data were calibrated with the {\sc MIRIAD} package (Sault et al.\
1995).  The continuum subtraction was done by using a linear fit through the
line-free channels of each visibility record and subtracting this fit from all
the frequency channels (``UVLIN'').  

The estimated rms noises given in Tables 2a, b and c are taken from Hanning
smoothed images obtained with natural weighting.
We have also used the line-free channels for producing an image of the
continuum emission. 

%
%%% Table 2
% 
\begin{table*}
{\sc Table 2a.} {Log of the observations: ATCA}
\smallskip

\begin{tabular}{cccccccc} \hline \hline
 Name   & Other    &   Date & Freq.  & Band & Exp.  & Beam (PA) & rms \\
        & name     &         & MHz   & MHz  &   Time(h) & $^{\prime\prime}\, \times 
^{\prime\prime}\, (^{\circ})$ & mJy/beam \\ \hline
0428--53 &          & 98/08/29 & 1367 & 16 & 3h   & 7.5$\times$6.3 (+30) & 3.34\\
0518--45 & Pic A     & 97/11/07 & 1372 & 16 & 3h   & 8.1$\times$5.8 (+8)  & 3.20\\
0521--36 &          & 98/08/29 & 1346 & 16 & 3.5h & 10.0$\times$6.8 (+8) & 3.15\\
0620--52 &          & 97/11/07 & 1349 & 16 & 3h   & 7.8$\times$6.4 (+83) & 2.40\rlap{$^*$}  \\
         &          & 98/08/30 & 1355 & 16 & 3h   &                &\\
0625--35 &          & 97/11/07 & 1349 & 16 & 3h   & 9.2$\times$5.2 (4.3) & 2.68\\
1246--41 & NGC~4696 &  97/11/08 & 1405 & 16 & 3.5h & 7.5$\times$7.0 (69) & 2.60\\
1318-43 & NGC~5090 &   97/11/08 & 1405 & 16 & 3.5h & 7.6$\times$5.5 (--4) & 1.55\rlap{$^*$}\\
        &          &   98/05/16 & 1405 & 16 & 10h &  & \\
1333-33 & IC~4296  &   97/11/08 & 1405 & 16 & 3.5h & 9.2$\times$6.3 (+3) & 2.79 \\
1549-79 &          &   97/11/08 & 1235 & 16 & 3h & & 5.67\\
1637-77 &          &   97/11/06 & 1365 & 16 & 3h & 6.5$\times$5.8 (27) & 2.69\\
1814-63 &          &   97/11/06 & 1336 & 16 & 3h & 5.7$\times$5.4 (62) & 5.25\\
2152-69 &          &   97/05/11 & 1382 & 16 & 10h & 6.7$\times$6.3 (--15) & 2.02\\
\hline
\end{tabular}

%\end{table*}
%\begin{table*}

{\sc Table 2b.} {Log of the observations: VLA}
\smallskip

\begin{tabular}{ccccccccc} \hline \hline
 Name   & Other    &   Date & Freq.  & Band & Exp.  & Beam (PA) & rms\\
        & name     &         & MHz   & MHz  & Time(min)    & $^{\prime\prime} \times 
\,^{\prime\prime}\, (^{\circ})$ & mJy/beam \\ \hline
0034--01 & 3C~15    & 98/09/29 & 1324 & 6.25  & 15  & 7.2$\times$6.3 (--2) & 0.65\rlap{$^*$}\\
         &         & 99/10/30 &      &       & 50  & &\\
         &         & 99/11/04 &      &       & 40  & &\\
0035--02 & 3C~17    & 99/10/30 & 1164 & 6.25  & 50  & 8.6$\times$5.7 (+46) & 2.80\\
         &         & 99/11/04 &      &       & 40  & &\\
0055--01 & 3C~29    & 98/09/29 & 1360 & 6.25  & 30  & 8.0$\times$6.4 (--36)& 1.00\\
0123--04 & 3C~40    & 99/10/30 & 1395 & 6.25  & 50  & 7.3$\times$5.3 (+44)& 0.46\rlap{$^*$} \\
         &         & 99/11/04 &      &       & 40  & &\\
0305+03  & 3C~78    & 98/09/30 & 1381 & 6.25  & 12  & 6.4$\times$5.5 (--54)& 0.95\\
0325+02  & 3C~88    & 98/09/30 & 1379 & 6.25  & 30  & 6.5$\times$4.8 (--56)& 0.83\\
1251--12 & 3C~278   & 98/09/29 & 1399 & 6.25  & 30  & 11.4$\times$5.7 (+35)& 1.50 \\
1333--33 & IC4296  & 98/09/29 & 1403 & 6.25  & 30  & 19.7$\times$4.9 (+22)& 1.90\\
1717--00 & 3C~353   & 98/09/29 & 1378 & 6.25  & 40  & 4.4$\times$3.7 (--81)& 0.80\rlap{$^*$}\\
         &         & 99/10/30 &      &       & 50  & &\\ 
1949+02  & 3C~403   & 98/09/29 & 1341 & 6.25  & 90  & 5.9$\times$5.2 (--77)& 0.65\\
2221--02 & 3C~445   & 98/09/29 & 1345 & 6.25  & 30  & 10.0$\times$6.5 (--44)& 0.92\rlap{$^*$}\\
         &         & 99/10/30 &      &       & 50  & &\\ 
2314+03  & 3C~459   & 99/10/30 & 1165 & 6.25  & 50  & 6.1$\times$4.3 (+67)& 3.82\\
\hline
\end{tabular}
%\end{table*}

%\begin{table*}
{\sc Table 2c.} {Log of the observations: WSRT}
\smallskip

\begin{tabular}{cccccccc} \hline \hline
 Name   & Other    &   Date & Freq.  & Band & Exp.  &  rms & Peak\\
        & name     &         & MHz   & MHz  & Time(min)      & mJy/beam & Jy\\ \hline
2314+03  & 3C~459   & 00/05/24 & 1164.5 & 10  & 3.6h  &  1.25 & 5.007\\
\hline
$*$ rms noise from the combined data
\end{tabular}
\end{table*}

A few more objects from the 2-Jy sample have data available from the
literature: NGC~4261 (PKS~1216+06, 3C~270) observed by Jaffe \& McNamara
(1994) and recently by van Langevelde et al.\ 2000; Hydra A observed by
Dwarakanath et al.\ (1995) and Taylor (1996); 3C~317 (PKS~1514+07) observed by
Jaffe (1991), Centaurus A (van der Hulst et al.\ 1983), PKS~1934--63
(V\'eron-Cetty et al.\ 2000), PKS~0023--26 and PKS~2135--20 (Pihlstr\"om et
al.\ in prep).  We shall consider also these objects in our discussion.

\section{Results}

Of the 23 radio galaxies we observed, five galaxies show \HI\ absorptions: PKS
1318--43 (NGC~5090), PKS 1549--79, PKS 1814--63, PKS 1717--00 (3C~353) and PKS
2314+03 (3C~459).  No \HI\ in emission has been detected.  The results
obtained for our sample are summarised in Table~3. $\Delta S$ represents the
maximum depth of the absorption, or its 3$\sigma$ limit.

The line profiles for the five detected objects are shown in Figs.  1 to 5.
The $x$-axis shows the heliocentric velocities derived using the optical
definition.
The systemic velocities (defined as $V_{\rm hel}=cz$) are marked
in the plots.  The values are taken from the literature (see the notes for
details), except in the case of PKS~1549--79 and PKS~1814--63.  For these
objects the velocities are derived from new or re-analysed spectra and more
details are given below in the notes.  For the extended
objects an image (obtained using the line-free data and using uniform
weighting) of the radio continuum is also shown.

%
%%% Table 3
%
\begin{center}
\begin{table*}
{\sc Table 3.} {Parameters of the \HI\ absorptions}
\smallskip

\begin{tabular}{ccccccc} \hline \hline
 Name  & Other      & $S_{\rm c, 1.4\,GHz}$  & $\Delta S$(Jy)  & $\tau$  & $W_{20}$ &
N$_{\rm HI}/T_{\rm spin}$  \\
       & name      &  Jy & & \%  &  km s$^{-1}$ &  
       $10^{18}$ cm$^{-2}$   \\ \hline
0034--01 &  3C 15     & 0.702     & \llap{$<$} 0.002 & \llap{$<$} 0.3  & ...    & \llap{$<$} 0.15\rlap{$^*$}     \\
0035--02 &  3C 17     & 0.981     & \llap{$<$} 0.012 & \llap{$<$} 0.9  & ...    & \llap{$<$} 0.47\rlap{$^*$}     \\
0055--01 &  3C 29     & 0.060     & \llap{$<$} 0.003 & \llap{$<$} 5.1  & ...    & \llap{$<$} 2.82\rlap{$^*$} \\
0123--04 &  3C 40     & 0.057     & \llap{$<$} 0.002 & \llap{$<$} 2.5  & ...    & \llap{$<$} 1.35\rlap{$^*$}     \\
0305+03  &  3C 78    & 0.716     & \llap{$<$} 0.004 & \llap{$<$} 0.4  & ...    & \llap{$<$} 0.22\rlap{$^*$}     \\
0325+02  &  3C 88    & 0.091 & \llap{$<$} 0.003 & \llap{$<$} 2.8  & ...    & \llap{$<$} 1.52\rlap{$^*$}     \\
0428--53 &           & 0.165 & \llap{$<$} 0.010 & \llap{$<$} 6.3  & ...    & \llap{$<$} 3.44\rlap{$^*$}     \\
0518--45 & Pic A      & 0.728 & \llap{$<$} 0.010 & \llap{$<$} 1.3  & ...   & \llap{$<$} 0.73\rlap{$^*$} \\
0521--36 &           & 5.546 & \llap{$<$} 0.009 & \llap{$<$} 0.2  & ...   & \llap{$<$} 0.09\rlap{$^*$}      \\
0620--52 &           & 0.368 & \llap{$<$} 0.007 & \llap{$<$} 2.0  & ...   & \llap{$<$} 1.08\rlap{$^*$}      \\
0625--35 &           & 1.200 & \llap{$<$} 0.008 & \llap{$<$} 0.7  & ...   & \llap{$<$} 0.37\rlap{$^*$} \\
1246--41 & NGC~4696  & 0.920 & \llap{$<$} 0.008 & \llap{$<$} 0.9  & ...   & \llap{$<$} 0.47\rlap{$^*$} \\
1251--12 &  3C 278    & 0.069 & \llap{$<$} 0.005 & \llap{$<$} 6.7  & ...   & \llap{$<$} 3.70\rlap{$^*$}      \\
1318--43  & NGC~5090 & 0.275 &  0.008   & ~2.9    & 150   &   2.7 \\
1333--33  & IC~4296  & 0.156 & \llap{$<$} 0.006 &  \llap{$<$} 4.1 & ...   & \llap{$<$} 2.23\rlap{$^*$}\\
1549--79  &          & 4.73  &   0.120   &   2.6  &  100-200    & 3.6  \\
1637--77  &          & 0.175 & \llap{$<$}0.008  & \llap{$<$} 4.7 & ...    & \llap{$<$} 2.59\rlap{$^*$}\\
1717--00  & 3C 353    & 0.094 &  0.010    & 10.9    & 300   & 42     \\
1814--63  &          & 11.9  &  2.280    & 21.3   & 75     & 17   \\
1814--63  &          & 11.9  &  0.100    & 0.8    & 300   & 20   \\
1949+02  & 3C 403    & 0.021 & \llap{$<$} 0.002      & \llap{$<$} 9.8 & ...    & \llap{$<$} 5.35\rlap{$^*$}     \\
2152--69  &          & 0.583 & \llap{$<$} 0.005      & \llap{$<$} 1.0 & ...    & \llap{$<$} 0.57\\
2221--02  & 3C 445    & 0.064 & \llap{$<$} 0.003      & \llap{$<$} 4.4 & ...    & \llap{$<$} 2.42  \\
2314+03\rlap{$^a$}  & 3C 459    & 3.932 &    0.033      &   0.8  & 400    & 2.7   \\
\hline
\hline

\end{tabular}
        
$^*$Values derived  using a width of 30 km s$^{-1}$ and 3$\sigma$ noise level. \\
$^a$ VLA data \\

\end{table*}
\end{center}

For all galaxies we have derived both the peak optical depth and the
integrated column density of the \HI\ (or the limit to that) and the results
are given in Table 3.  A more detailed description of the detected objects is
given in the notes to the sources.  The measured depth of the \HI\ absorption
line, $\Delta S$, depends on both the optical depth $\tau$ and the covering
factor $c_{\rm f}$ as: $\Delta S = Sc_{\rm f}(1-e^{-\tau})$ where S is the
continuum flux density.  The peak optical depths (column 6 in Table 3), $\tau
= -\ln (1- \Delta S/(S c_{\rm f}))$, are estimated using the flux (core or
total depending on the size of the source) derived from the continuum image
obtained from the line-free channels in our observations (see column 4 of
Table 3 for the flux density of the nucleus).  For our calculations the
covering factor has been assumed to be $c_{\rm f}=1$ and we will not be able
to put constrains on this value until we can image the absorption using VLBI. 
Thus, the derived optical depths (and the column densities) can be considered
as lower limits.  The values of the peak optical depth go from 0.5\% for
3C~459 and $\sim $3\% for PKS~1318--43 and PKS~1549--79, up to $\sim$11\% for
3C~353 and $\sim$21\% for PKS~1814--63.  The full width of the \HI\ profiles
measured at 20\% of the peak profile intensity ($W_{20}$) are given in column
7 of Table 3.

The integrated column density (column 8 in Table 3) has been
estimated as: $N_{\rm H} \simeq 1.83\times 10^{18} T_{\rm spin} \int \tau
\,dv$, where \tspin\ is the spin temperature in Kelvin. 

The spin temperature is a very uncertain parameter in deriving the column
densities.  The value of the spin temperature is essentially unknown for
extragalactic sources, so typical Galactic values (between 100 and 1000 K,
Heiles \& Kulkarni 1988) are often assumed.  However, the presence of a strong
continuum source near the \HI\ gas may significantly increase the spin
temperature because the radiative excitation of the \HI\ hyperfine state can
dominate the usually more important collisional excitation (see e.g.  Bahcall
\& Ekers 1969).  If the gas producing the \HI\ absorption is in a
circumnuclear torus (i.e.  at a distance $\leq 100$ pc from the nucleus), it
will be irradiated by the hard X-ray emission from the nucleus.  According to
Maloney et al.  (1996, and references therein), purely atomic gas in these
conditions has a stable equilibrium temperature of $\sim $8000 K. 

For the undetected galaxies, a 3$\sigma$ limit to the peak optical depth was
estimated and is given in Table~3.  For these galaxies, the upper limits to
the column density (also given in Table 3) were calculated by assuming a
profile width of 30 \kms.  The upper limits to the optical depth are typically
between 0.5 and a few \%, with the extreme values of $\sim $9\% for 3C~278 and
3C~403 -- objects with particularly weak cores.  Some of the upper limits are
due to observations not being deep enough, but many are significant upper
limits indicating that in those objects no \HI\ absorption is observed at the
level at which it is typically observed in the other radio galaxies.

Below we give some more details about the objects in the sample detected in \HI\
absorption. 

\subsection{PKS~1318--43 (NGC~5090)}

% Fig. 1a
\begin{figure}
\centerline{\psfig{figure=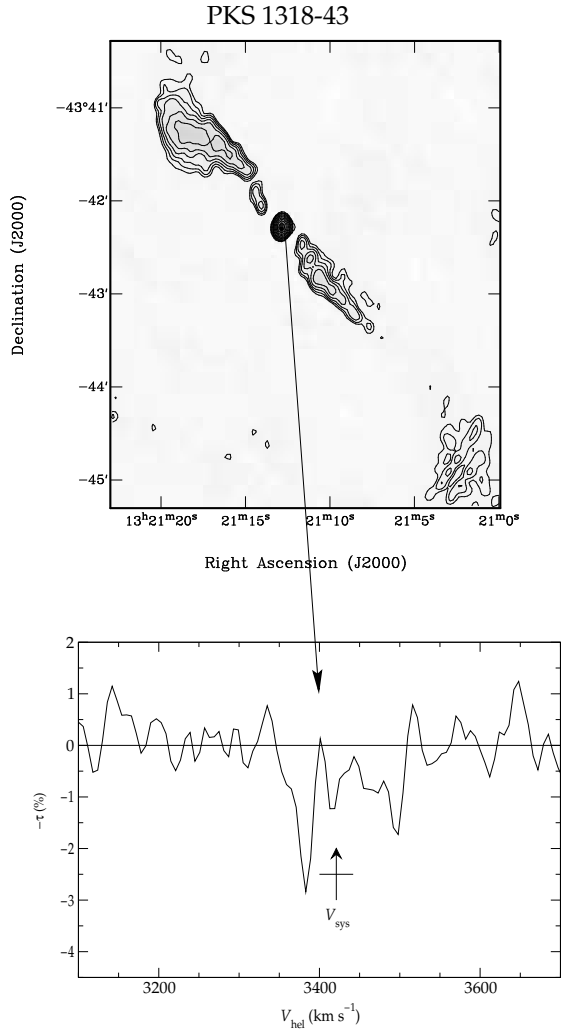,width=8cm,angle=0}}
\caption{Radio continuum image ({\sl Top}) of PKS~1318--43 (NGC~5090) at 21 cm
obtained from the line-free channels. The contour levels are: 3.5 mJy
beam$^{-1}$ to  274 mJy beam$^{-1}$ in steps of factor 1.5.
\figHI\ absorption spectra ({\sl Bottom}) observed
against the core of the radio galaxy PKS~1318--43 (NGC~5090). The systemic
velocity (from Tadhunter et al. 1993) is indicated.}
\end{figure}

This is a typical FR-I radio galaxy as can be seen from the radio images
presented in Morganti et al.  (1993) and Lloyd, Jones \& Haynes (1996).  In
Fig.~1 we present the continuum image obtained from our data as described
above.  NGC~5090 is believed to be part of an interacting system together with
a nearby spiral galaxy NGC~5091 (Smith \& Bicknell 1986). 

\HI\ absorption is  detected against the core (see Fig.~1).  The \HI\ 
profile appears double peaked, although the relatively high noise level
does not allows us to say much about its shape.  The full width at 20\% of the
peak is $\sim $150 km/s.  The \HI\ velocity derived as arithmetic mean of the
velocities at 20\% of the peak intensity of the \HI\ profile is $\sim$3425
\kms.  This is very close to the systemic velocity of the galaxy (3421 \kms, 
Tadhunter et al. 1993).  The peak optical depth is about 3\% and the
integrated column density $2.7 \times 10^{18} T_{\rm spin}$ cm$^{-2}$.

\subsection{PKS~1549--79}

PKS~1549--79 is a core-jet radio source of $\sim $400 pc (about 120 mas) in
size and has been studied in detail at radio frequencies by King (1994) and
King et al.  (1996) using the southern VLBI network.  Overall it has quite a
flat spectral index because the flux is dominated by the core component (about
3~Jy at 2.3 GHz and  spectral index  $\alpha \sim 0.0$).

% Fig. 2
\begin{figure}
\centerline{\psfig{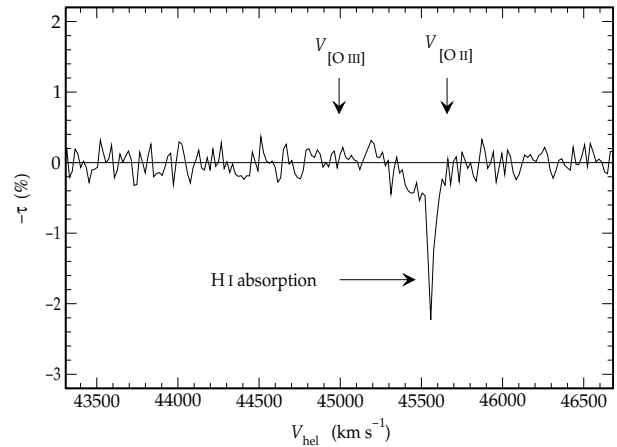}}
\caption{\figHI\ absorption spectra (after Hanning smoothing) in the source
PKS~1549--79. 
Two different velocities have been derived from the \oiii\ and \oii\
emission lines. See text for details.} 
\end{figure}

The \HI\ absorption detected (see Fig.\ 2) appears to be formed by a narrow
component ($W_{20} \sim$ 100 \kms) peaked at $\sim $45580 \kms, although a
broader component ($W_{20}=$ 250 \kms), mainly on the blue side of the narrow
component, is possibly detected.  The peak optical depth of the narrow
component is about 2.5\%.  The
integrated column density (including the broad absorption) is $\sim $$3.6 \times
10^{18} T_{\rm spin}$ cm$^{-2}$.

A new optical spectrum was obtained for this galaxy with the 3.6-m ESO
telescope.  A detailed discussion of the optical characteristics of
PKS~1549-79 will be presented in Tadhunter et al. (2000).  For the
interpretation of the \HI\ absorption the finding that the low and high
ionization lines (\oii\ and \oiii\ respectively) give quite different values
for the redshift is particularly relevant.  From the \oiii\ line a value of
$z=0.15008 \pm 0.00016$ (or $V_{\rm hel} = 44993 \pm 48$ \kms) is derived,
while from the \oii\ line we obtain $z=0.1523 \pm 0.0002$ (or $V_{\rm hel} = 45658
\pm 60$ \kms).  An other interesting characteristic is that the \oiii\ line has a
width of $\sim $1345 \kms\ FWHM.  This  is one of the largest widths measured
in radio galaxies.  The \oii\ has a smaller width (about 650 \kms\ FWHM).
Thus, we believe that much of the \oiii\ is emitted by an inner narrow line
region, which is undergoing {\sl outflow} from the nucleus, whereas the
\oii\ emission  is likely to come from  a outer region (less affected by the
outflow).  This result is further  discussed in \S 5.

An other interesting feature  of this galaxy is that indications of ongoing
star-formation have been found by  carefully fitting the optical
spectrum (Dickson 1997).  This is quite rare for powerful radio galaxies. 
PKS~1549--79 is also strong in the IR: despite the high redshift it has been
detected by IRAS with a
IR luminosity of $L_{\rm IR} = 1.6 \times 10^{11}$ \Lsun\ if we use the
100$\mu \rm m$ upper limit as a detection.  This derived luminosity is high
for a radio galaxy.

% Fig. 3a
\begin{figure}
\centerline{\psfig{figure=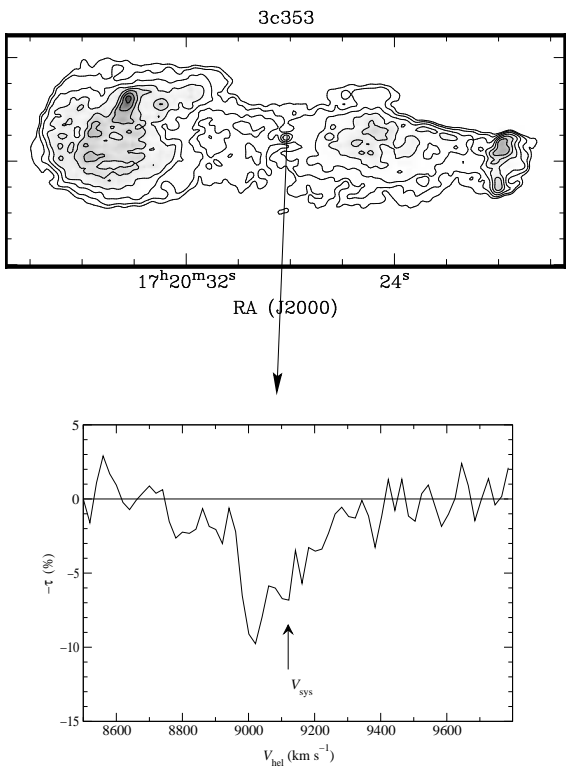,width=8cm,angle=0}}
\caption
{Radio continuum image ({\sl Top}) of 3C353 obtained from the line-free
  channels.  The contour levels are: 20 mJy beam$^{-1}$ to 417 mJy beam$^{-1}$
  in steps of factor 1.5.\figHI\ absorption spectra ({\sl Bottom}) observed in
  the source 3C~353 (PKS 1717--00). The cross represent the systemic velocity
  (see text)}
\end{figure}

\subsection{PKS~1717--00 (3C 353)}

PKS~1717--00 is a strong radio galaxy with a FR-II morphology.  The radio
structure is shown in the continuum image in Fig.3 and a detailed study of the
structure of this source, and in particular of the well collimated jet (and
counter-jet), has been done by Swain, Bridle \& Baum (1999).  Despite being a
powerful radio galaxy, only weak \oiii\ and H$\alpha$ emission lines have been
detected in the spectrum of this object (Tadhunter et al. 1993).  The optical
continuum of this object is very red (Simpson et al. 1996).

At the position of the radio core we detect \HI\ absorption (see Fig.
3) with an optical depth of $\sim $11\% and a width of $\sim 300$ \kms (but
this  is very uncertain due to the poor signal-to-noise).  The integrated column
density is quite high: $\sim $$4.2 \times 10^{19} T_{\rm spin}$  cm$^{-2}$

The systemic velocity is $9120 \pm 59$ \kms\ (or $z = 0.030421 \pm 0.000197$)
from the RC3 (De Vaucouleurs et al. 1991), very close to the peak of the \HI\
absorption.

\subsection{PKS~1814--63}

PKS~1814--63 is a Compact Steep Spectrum (CSS) source with a basic double-lobed
structure oriented almost  north-south (Tzioumis et al.  1996).  Because VLBI
radio continuum observations are available for this galaxy at only one
frequency (2.3 GHz), no information is available on which of these structures
(if any) correspond to the core.  The overall extent of the source is 410 mas
corresponding to 328 pc.

% Fig. 4
\begin{figure}
\centerline{\psfig{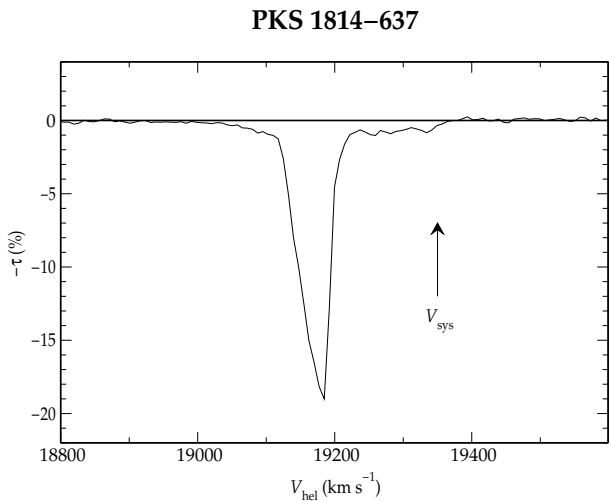}}
\caption{\figHI\ absorption spectra (after smoothing) in the source
PKS~1814-63. The systemic velocity recently determined is indicated (see text
for details).} 
\end{figure}

In PKS~1814--63 deep \HI\ absorption is detected. V\'eron-Cetty et al. (2000)
have also independently detected \HI\ absorption in this galaxy.  The profile
can be seen in Fig.4 and it shows deep absorption together with a broad
shallow feature.  The former component is centred on $V_{\rm hel} = 19180$
\kms\ (corresponding to $z \sim 0.06393$), has $W_{20} \sim$ 75 \kms\ and a
peak optical depth of $\sim $20\%.  The broad component has $W_{20} \sim 300$
\kms\ centered on $V_{\rm hel}=19220$ \kms, and has an optical depth of a few
percent, more similar to what usually found in other galaxies.  The integrated
column density of the deep component is $\sim $$1.7 \times 10^{19} T_{\rm
  spin}$ cm$^{-2}$.  Including also the broad component the column density is
$\sim 2 \times 10^{19} T_{\rm spin}$ cm$^{-2}$.

We have re-analyzed the optical spectrum presented in Tadhunter et al.
(1993).  Because of the presence of a bright star near the object, the quality
of the spectrum is low.  However, by doing a more careful extraction of
the spectrum, we have now detected the emission line at higher S/N, and,
consequently, have been able to determine a more accurate redshift.
We derive a new value for the
redshift of the galaxy of $z = 0.0645$, corresponding to $V_{\rm hel}= 19350$
\kms.  This is different from the previously derived redshift.   Although
this new redshift will require further confirmation  with better data, the
\HI\ absorption appears now mostly blueshifted respect to the new systemic
velocity.

\subsection{PKS~2314+03 (3C~459)}

In this galaxy the optical continuum spectrum is bluer than that of a typical
early-type galaxy, and absorption features suggestive of young stellar
populations have been detected (Miller 1981).  The emission-line spectrum indicates moderate
ionization.

The radio morphology shows a strong nuclear component and two  very
asymmetric lobes (Ulvestad 1985 and Morganti et al.\ 1999).  The size of the radio
emission is only $\sim $10 arcsec (corresponding to $\sim $48 kpc).  The
central component detected in the VLA image (see e.g.\ Morganti et al.\ 1999)
has been found by higher resolution Merlin observations (P.\ Thomasson private
communication) to be resolved in two components, both with a steep spectral index.

% Fig. 5a
\begin{figure}
\centerline{\psfig{figure=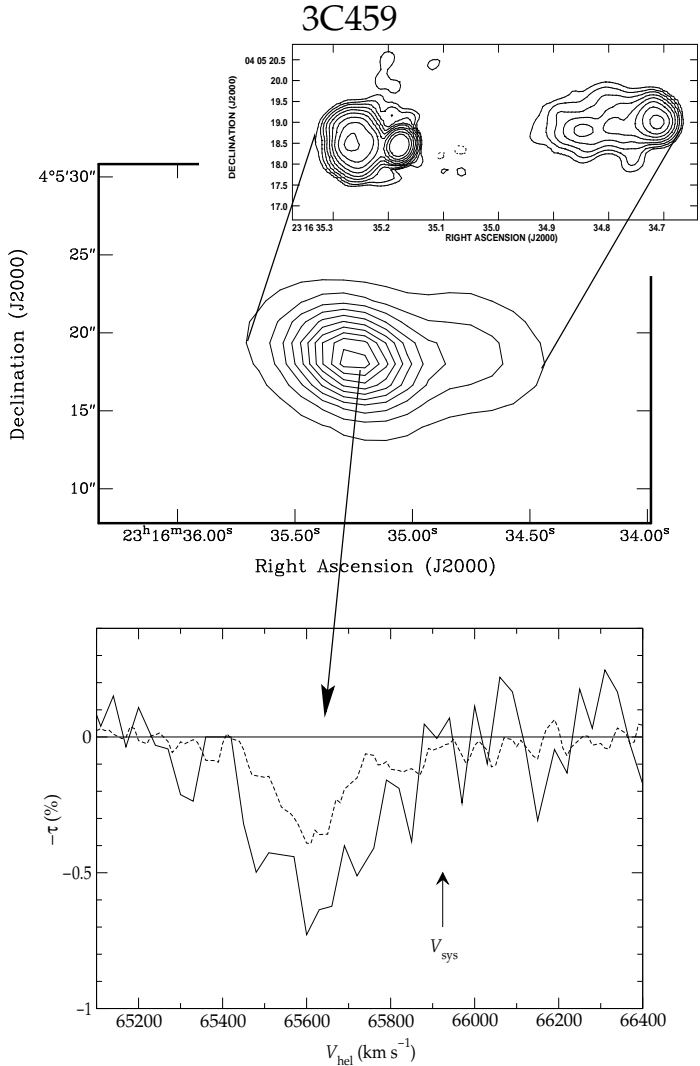,width=10cm,angle=0}}
\caption
{Radio continuum image of 3C459 from VLA at 6cm ({\sl Top pannel}, from
Morganti et al.  1999) and from the line free channels of the VLA 21 cm data
presented in this paper ({\sl Middle pannel}, the contour levels are from
0.195 to 3.9 Jy beam$^{-1}$ in steps of 390 mJy beam$^{-1}$.  {\sl Bottom pannel}:
\figHI\ absorption spectra (from the VLA data (after smoothing, fill line)
and from the WSRT data (dashed line).  The systemic velocity (from Spinrad et
al.\ 1985) is indicated.} \end{figure}

At the resolution of the WSRT data the continuum emission from 3C~459 is
unresolved, while at the resolution of the VLA the continuum emission is 
slightly resolved  (see Fig.\ 5).

The peak optical depth is only about 0.7\% in the VLA data and even lower
($\sim $0.4\%) in the WSRT data. The difference between the two is most likely
due to a dilution effect: the WSRT data include the continuum flux from the entire
source while from the VLA data it is clear that the absorption is coming from
the core and/or eastern lobe. Only higher resolution observations will tell us
whether the absorption is really against the core or more against the
lobe. The FWHM of the absorption appears to be quite broad in our VLA data
($\sim $400 \kms) and is confirmed by the higher quality WSRT profile shown in
Fig. 5. The absorption is centered on $V_{\rm hel} \sim 65550$ \kms.  The systemic
velocity is 65924 \kms\ (corresponding to a redshift of 0.2199, Spinrad et al.
1985).

The integrated column density derived from the VLA data is similar to those
found for most other radio  galaxies: $\sim 2.7 \times 10^{18} T_{\rm spin}$
cm$^{-2}$.

\section{Occurrence of \HI\ absorption and the relationship with other
characteristics}

\HI\ absorption is detected in 5 galaxies of the 23 observed.  The
detection rate is between 20 and 30\% depending whether we also consider the
data available in the literature.  Among the latter, 7 galaxies were observed and 5
detected: these numbers are likely to be biased toward detection (it is less
likely that non-detections are published!).  A similar detection rate was
found in the only other systematic search for \HI\ absorption (van Gorkom et
al.\ 1989).

We can now investigate how the detection rate is distributed among galaxies
with different optical and radio properties.

\begin{enumerate}

\item Of the {\sl 10 Fanaroff-Riley type-I radio sources observed (including also the intermediate
galaxy 3C~15), we detect absorption in only one galaxy} (PKS~1318--43).  A few
more cases can be included if the data from the literature are considered (e.g.
Hydra~A, Centaurus~A and NGC~4261).  Nevertheless, the fraction of detections
remains small.

\item So far, {\sl no \slHI\ absorption has been detected in
any broad line radio galaxy} in our sample (see also below).

\item In both {\sl compact powerful radio galaxies (PKS~1814--63, PKS~1549--79)
observed by us, \slHI\ absorption is detected}.  3C~459 is also a relatively
small object and has a compact steep spectrum core (that has been interpreted
as an indication of objects embedded in gas-rich environments, van Breugel et
al. 1984) and it also shows \HI\ absorption.  Two more compact sources in the sample
were observed with the WSRT (Pihlstr\"om et al. in prep.).  Only in one of the two
\HI\ absorption was detected (PKS~0023--26).  However, it is worth mentioning
that the undetected source (PKS~2135--20) is likely to be a broad line radio
galaxy.  Broad wings have been detected in the Mg{\,\small II} line of this
galaxy (Shaw et al. 1995).  This would be consistent
with the trend for BLRGs described above.

\item In the recently completed spectroscopic survey of 20 southern 2-Jy
radio galaxies ($0.15 < z < 0.7$, Tadhunter et al. in prep.) only two galaxies
(3C 459 and PKS~1549--79) are clearly dominated by starbursts at optical/UV
wavelengths, and these two objects are also the only two in our sample
to be detected by IRAS at 60 $\mu$m.  It is remarkable that {\sl \slHI\
absorption is detected in both of these galaxies}.

\item \HI\ absorption is detected in a FR-II radio galaxy  with weak
emission lines: PKS~1717--00 (3C~353).  Although the upper limits for the other
two galaxies of this kind (3C~88 and PKS~1637--77) are quite high, they do not
show \HI\ absorption at the  level detected in PKS~1717--00. 

\item no \HI\ absorption is detected in the BL Lac object, PKS 0521--36.

\end{enumerate}
 
A discussion of the  origin of the \HI\ absorption will be made in \S 6.

\section{The problem of the systemic velocity}

An important issue for understanding the origin of the \HI\ absorption is how
the systemic velocity of the galaxy compares with the velocity of the \HI. 
This can be a complicated (and often underestimated) problem and therefore it
is worthwhile to emphasize the uncertainties connected with the values of the
systemic velocities.  It was already pointed out by Mirabel (1989) how the
systemic velocities derived from emission lines can be both uncertain and
biased by motions of  the emitting gas.  As discussed in the notes
above, two of our galaxies detected in \HI\ illustrate these problems. 

In PKS~1549--79 the old redshift available in the literature (Tadhunter et al. 
1993) and derived from \oiii\ emission lines (as for the majority of radio
galaxies) suggested the \HI\ being strongly redshifted.  However, as discussed
above, from a newly obtained optical spectrum we find that the \oii\ lines
give a radial velocity about 660 \kms\ higher compared to that derived from
the \oiii\ line, and a velocity more {\sl consistent with the \slHI\
absorption} (see Fig.~1).  What should we consider as systemic velocity of the
galaxy? The \oiii\ line appears to be very broad, one of the broadest observed
in radio galaxies, suggesting that this line comes from a region where the gas
is very disturbed.  This region could be closer to the nucleus, in agreement
with the suggestion made by Hes et al. (1993), while the \oii\ lines would originate
farther away from the nucleus and from a less disturbed region.  Thus, the
value derived from the \oii\ lines is likely to be closer to the real systemic
velocity of this galaxy. 

PKS~1549--79 is not the only object in which this effect is observed. Grandi (1977)
pointed out that in  4C12.50 (PKS~1345+12) two redshift systems were
found: one containing the high-ionization lines and the other (at +240 \kms\
with respect to the first one) containing the low-ionization lines. Moreover,
although a large uncertainty is present, the redshift of the \HI\ absorption
features appears to be the same as that of the low-ionization emission
lines, therefore supporting this as the systemic velocity.  We will come back
to the similarities between PKS~1549--79 and 4C 12.50 in \S 6.2.3.

In the case of the other compact object (PKS~1814--637), a more accurate
redshift has been derived by better separating the emission from the AGN from
the nearby star.  The \HI\ absorption appears now to be mainly blue-shifted
with only part of the shallow absorption observed at the systemic velocity
(unlike what was obtained using the old value for the systemic velocity) as shown
in Fig.~4.

In summary, the results we have obtained from our attempt to derive a more
detailed value for the redshift suggest that:

1) the redshift derived from the \oiii\ line can sometimes be unreliable.  The
\oiii\ emission line can be more affected by outflow than low ionization lines
like \oii\ because the former may be coming from an inner region, closer to
AGN, in agreement with what was claimed by Hes et al. (1993).

2) some of the redshifts available in the literature are too uncertain for a
meaningful comparison with the \HI\ absorption and in particular to derive
reliable conclusions on the motion of the neutral gas with respect to the
galaxy. 

In the case of our detected objects (and with the redshift available so far)
we find that in two cases (PKS 1318--43 and PKS 1717--00) the \HI\ absorption
is at the systemic velocity and in two objects (PKS 1814--63 and 2314+03) all or
most of the \HI\ absorption is blueshifted compared to the systemic.

However, better quality spectra will be necessary to confirm this result in
particular for 2314+03.  Finally, in the case of PKS 1549--79, the \HI\ 
absorption has a velocity about 100 \kms\ lower than that derived from the
\oii, therefore, considering the errors, is only marginally blueshifted.

\section{The origin of the \HI\ absorption}

As mentioned in the Introduction, circumnuclear disks or tori are believed to
be the most likely causes of \HI\ absorption in radio galaxies.  Thick or
warped disks surrounding the central regions are required by unified schemes
in the case of FR-II radio galaxies to explain the large column densities seen
in X-ray observations, the broad lines seen in reflected light and the far-infrared
colours.  On the other hand, while the unification of BL Lacs and FR-I
requires relativistic beaming, the presence of such a thick disk in such
objects is still a
matter of discussion (Urry \& Padovani 1995).  In particular, while in BL Lacs
broad lines have been detected (Vermeulen et al.  1995, Corbett et al.  1996)
the situation is still unclear for FR-Is.  Thus, for these radio galaxies the
presence of thick disks surrounding the central black hole is not 
required to explain the existing observations in term of the unified schemes.

Because of this possible difference between the two types of radio galaxies,
let us consider them separately.

\subsection {Thin nuclear disks in low luminosity radio galaxies?}

We have detected \HI\ absorption in only one of the FR-I radio galaxies of the
23 sources observed.  The absorption in this case appears to be at the
systemic velocity with a width of $\sim $150 \kms
and a relatively low column
density (2.7 $\times 10^{18} T_{\rm spin}$ cm$^{-2}$).  The absorption line
could be due to a nuclear disk as is suggested for some of the other FR-I
galaxies of the 2-Jy available the in literature (e.g.  Hydra A, NGC 4261).
This is in contrast with the narrow profiles observed in galaxies like
Centaurus~A which were interpreted in term of single clouds in front of the
radio continuum.

At a first glance, the low detection rate of \HI\ absorption in FR-I galaxies
is somewhat surprising given that, as shown from recent HST images, gas and
dust, in particular circumnuclear dust-lanes, are commonly present in
low-luminosity radio galaxies (see e.g. de Koff et al. 1996, Chiaberge et al.
1999, Capetti et al. 2000).

\begin{figure}
\centerline{\psfig{figure=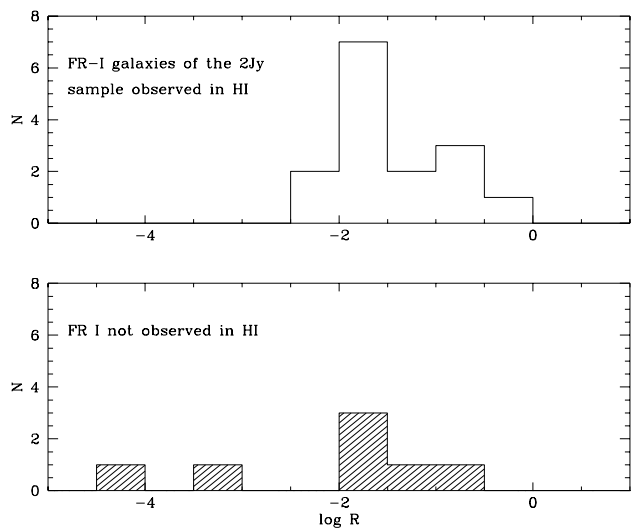,width=10cm,angle=0}}
\caption{Distribution of $R$ (ratio of core to extended 
radio fluxes, at 4.8 GHz from Morganti et al. 1993) for ({\sl Top}) the FR-I
radio galaxies of the 2Jy sample observed in \HI\ (either from this work or
from the literature)  and ({\sl Bottom}) for the
remaining FR-I galaxies in the 2Jy sample.} 
\end{figure}

One possibility to explain the low detection rate is that the selection of the
objects is biased: because only the objects with a relatively strong core
could be observed, we may have biased the sample to objects more beamed toward
us.  This is very likely to be the case for some objects: for example in the
case of 3C 78, the image taken by the Hubble Space Telescope (HST) shows
point-like nuclear emission (Chiaberge et al.  1999) and there is evidence for
the presence of an un-absorbed, nuclear X-ray source, (Trussoni et al.  1999)
both supporting the lack of nuclear obscuration as related to the particular
viewing angle.  Similar conclusions have been drawn from X-ray observations of
PKS~0625--35 (Trussoni et al.  1999).

To check how much such a bias may have affected our sample, we have compared
the distribution of $R$ for the FR-I radio galaxies of the 2Jy sample observed
in \HI\ (either from this work or from the literature) with the values of $R$
for the remaining FR-I galaxies (i.e. not yet observed in \HI).

The two distributions are presented in Fig.~6.  The values of $R$ of the two
distributions cover a similar range, with only two cases (Fornax~A and
Hercules~A) of very low values of $R$ among the objects not observed in \HI. 
Indeed, they were not observed because of the very low values of their core
flux.  Thus, the galaxies observed in \HI\ do not appear to be strongly biased
toward high $R$ values.  Moreover, we note that Hydra A, one of the FR-I in
the sample with a relatively low value of $R$ ($ \log R= -2.17$), shows \HI\
absorption at $\sim$50\% level (Dwarakanath et al.\ 1995) and even higher in
the VLBI observations while other objects with similar value of $R$ are
undetected at the $\sim $1\% level. 

Another possibility to explain the lack of \HI\ absorption is therefore to
assume that the nuclear disks, if present in these objects, are {\sl thin}.  The
presence of a thin disk has been claimed, e.g., in the case of NGC~4261 (Jaffe
\& McNamara 1994, van Langevelde et al.  2000).  For this object, the VLBI
data suggest that the \HI\ absorption is due to a disk of $\sim $1.3 pc
high projected against the counter-jet. 

Moreover, despite the presence of the dusty structures, optical studies show
that the nuclei of FR-I galaxies appear basically unobscured.  Chiaberge et
al.  (1999) find Central Compact Cores (CCC) in 85 \% of the FR-I galaxies
selected from the 3CR catalogue and observed with HST.  Thus, {\sl their
fraction of unobscured objects is comparable to the fraction of galaxies
undetected in \HI\ absorption}.  Based on the CCC detection rate, they derive
a thickness over size ratio $\ltae 0.15$.  If we take the values derived for
NGC~4261 (a circular velocity $V_{\rm c}$ of 610 \kms\ for a radius of $r =$ 5
pc and velocity dispersion $\sigma= 130$ \kms) and using the thin disk
relation (i.e.  the thickness of the disk is $h = r \sigma/ V_{\rm c}r $) we
find that this gives a 10\% probability of \HI\ absorption from the disk
against the nucleus, thus similar to the detection rate in the FRI sample as a
whole.

In summary, our results for FR-I galaxies support the idea the most of the
\HI\ absorption in these objects comes from a {\sl thin} nuclear disk as is
found for other galaxies of this type (NGC 4261 and Hydra A).  However, they
do not provide the definitive evidence against a thick torus because, e.g.  a
thick pc scale torus could be molecular especially given that the quasar
component in FRIs, if present, may be weak, so that there may be less
photo-dissociated molecular gas.  Nevertheless, the low detection rate of \HI\
absorption is consistent with that derived from optical work (Chiaberge et
al., 1999), i.e.  the ``standard'' pc scale, geometrically thick torus is not
present in low-luminosity radio galaxies. 

\subsection {Powerful radio galaxies: thick disks \& effects of the ISM?}

For FR-II radio galaxies, the presence of a {\sl thick} nuclear disk is a
vital ingredient in the unified schemes hypothesis.  Broad line radio galaxies
are supposed to be seen pole-on and therefore obscuration from the torus
should not occur.  Narrow-line radio galaxies (NLRG) are viewed more edge-on,
and it might be expected that \HI\ absorption would be more common in such
objects.  We will discuss this in some more detail. 

\subsubsection{Broad Line radio galaxies}

\HI\ absorption seems indeed absent in BLRGs, although the statistics are still 
limited.  For 3C~445 we only have a relatively high upper limit to the optical
depth: $\sim $4\%.  However, the remaining three have a tighter upper limit to
the optical depth of $\sim $1\%.  An even lower limit of $\sim $0.5\% has been
derived in the compact radio galaxy PKS~2135-20 (Pihlstr\"om et al. in prep.):
this is in clear contrast with the detections found for the other compact
sources in the sample. It appears that our data are consistent with the
assumptions that BLRGs are viewed pole-on and the absorption is related to an
obscuring torus.

The limiting column density derived for these objects ranges from $2 \times
10^{18} T_{\rm spin}$ cm$^{-2}$ down to $5 \times 10^{17} T_{\rm spin}$
cm$^{-2}$ .  Regardless of the value of $T_{\rm spin}$ assumed, the derived
column densities for the BLRGs are quite low.  These low column densities are
consistent with those derived from X-ray observations in the
case of the two BLRGs (PKS 2152--69 and Pictor A) for which X-ray observations
have shown that no soft X-ray absorption is present (Padovani et al.  1999).
On the other hand, in 3C~445 soft X-ray absorption has been detected
indicating a column density of $\sim 10^{23}$ cm$^{-2}$ (Sambruna et al.
1998).  The upper limit to the column density derived from the \HI\ is $< 2
\times 10^{18} T_{\rm spin}$ cm$^{-2}$ and even assuming a spin temperature of
$8000$ K, as discussed in \S 3, the column density will remain lower than the
one derived from the X-ray data. A similar discrepancy has been found between
the column densities derived from IR/optical data and X-ray data, with the
former much lower than the latter.  

In this case, the discrepancy between these values can be explained by
assuming that the radio-emitting jets in 3C445 are more spatially extended
than the region emitting the X-rays. Therefore they are less likely to be
affected by
HI absorption associated with dense material close to the central AGN.

\subsubsection{Narrow line radio galaxies}

Of the four narrow line radio galaxies observed by us, all but one are
detected and the undetected (3C~403) has a very high limit to the optical
depth due to the weakness of the radio core. 
The column densities found in our sample go from $\sim 3
\times 10^{18} T_{\rm spin}$ cm$^{-2}$ for PKS~1547--79 and 3C~459 up to a few
times $10^{19}$ cm$^{-2}$ for PKS~1814--63.  The latter value is similar to the
value found in the  prototype FR-II radio galaxy  Cygnus~A (Conway
\& Blanco 1995).  Two more objects (PKS 0023--26 and PKS 1934-63) have been
observed by other groups (see \S 2) and both detected.  The higher detection
rate in NLRGs compared to that in BLRGs is what is expected for unified
schemes. However, other factors may also play a role.

Due to the selection criteria, {\sl the NLRGs observed in our sample are mainly
compact or small objects}.  As pointed out by Heckman et al.  (1983), a radio
source of sub- or at most galactic size represents an optimum background
against which galactic \HI\ might be seen in absorption, in particular if the
radio continuum is confined to a scale comparable to the likely dimensions of
a possible nuclear disk.  Indeed, Conway (1996) found a high detection rate of
\HI\ absorption in compact objects.  Also from this study we confirm that
compact radio sources with narrow emission lines have a high chance of being
detected.  Of the 6 compact/small radio galaxies in the sample (including the
two galaxies observed by Pihlstr\"om et al.), the only one undetected is
likely to be a broad line radio galaxy (PKS 2135--20, Pihlstr\"om et al.).

Let us look in more detail to the three objects detected in this study. 
3C~459 shows a broad profile, at the resolution of our observations, but we
cannot be completely sure whether the absorption is against the nuclear region
or the eastern lobe.  Moreover, even assuming that the absorption is against
the former, it appears that the arcsec nuclear region is actually formed by
two steep spectrum components.  Thus, VLBI observations will be required to
clarify the real structure of the absorption.  The absorption is not at the
systemic velocity (although a more accurated redshift will be required to
confirm this)  but appears to be blueshifted.  Thus the torus, if present,
is misaligned with respect to the radio axis (and this would explain why
velocities different from the systemic are projectd against the core).  An
alternative explanation is that an outflow of the neutral gas is present.

The situation in PKS 1814--63 is similar. Although the double continuum
structure in PKS 1814--63 is larger than that usually observed in Compact
Symmetric Object (CSO), the morphology is very similar.  \HI\ absorption is
very often detected is CSOs (see e.g.  Peck \& Taylor 2000).  The likely size
of the atomic torus observed in other objects has a thickness between 50 and
100 pc at a distance of about 50-100 pc from the centre.  So if the absorption
is due to such a torus, it should appear only against one of the component.
Preliminary VLBI results (Morganti et al. 2000b) show that this is not the
case and \HI\ absorption is actually detected against both lobes (or part of
them).  Moreover, only part of the absorption, i.e.  part of the shallow
component, is at the systemic velocity (assuming the new redshift reported in
\S 3.4).  Most of the \HI\ absorption, including the very deep component,
appears to be blueshifted.  Thus, unless again the geometry of the disk
respect to the radio axis is peculiar, or
the systemic redshift substantially in error,
this part of the absorption will not be
easy to reconcile with a torus and could be instead gas associated with an
expanding gas cocoon around the radio source.

PKS 1549--79 seems to be a different kind of object.  This galaxy has a
core-jet radio structure and no counter-jet is observed at the resolution of
the VLBI ($\leq 10$ mas, $\sim 35$ pc, King et al.).  The detected \HI\
absorption has a lower column density and a narrower profile compared to the
other two galaxies.  With its core/jet VLBI morphology, it is difficult to
reconcile the absorption as due to a torus: in that case the absorption would
be expected against the counter-jet.  The characteristics of the \HI\
absorption in this object are more easy to reconcile with a cloud of neutral
hydrogen in front of the nucleus.

In summary, on one hand the fact that the \HI\ absorption is detected in NLRGs
and not in the BLRGs seems to suggest that a thick disk is indeed the cause of
absorption.  However, the above analysis shows that such a model does not
explain all the features of the \HI\ absorption in all NLRGs. The blueshifted
absorption does indicate outflow and the possible correlation of the detection
of \HI\ absorption with the presence of a particularly rich ISM (see below)
suggests that in some cases the absorption is related to an interaction
between the radio plasma and the ISM.  It is conceivable that this is
connected to the fact that all the detected sources are compact or small.
\HI\ absorption related to gas outflow has been clearly seen already in other
objects such as the Seyfert galaxy IC~5063 (Oosterloo et al. 2000), possibly the
superluminal source 3C 216 (Pihlstr\"om et al. 1999) and 3C 326 (Conway et al.
in prep).  Thus, it is not unlikely that a similar situations can be present
in some of our detected objects.

\subsubsection{Starburst and far-IR bright galaxies}

As mentioned in \S 4, only two galaxies (3C 459 and PKS~1549--79) out of the
20 powerful and intermediate redshift ($0.15<z<0.7$) radio galaxies of the
2-Jy sample are dominated by a starburst at optical/UV wavelengths (Tadhunter
1999) and these two objects are the only two in the sample detected by IRAS at
60$\mu$m.  It is therefore very interesting to see that in both these objects
\HI\ absorption has been detected.

Although a contribution to the far-IR emission from quasar heating cannot be
completely excluded, it is generally assumed that the far-IR excess represents
reprocessed starlight.  This points to a strong link between the star formation
activity and the far-IR, and to an overall richer interstellar medium present
in these galaxies.

In order to test if the rich ISM could have some correlation with detection of
\HI\ absorption, we have searched the literature to investigate if this trend
would hold for a larger number of objects.  As far as we could find, three
other starburst/far-IR bright galaxies were observed in \HI\ and {\sl all are
detected in absorption}.  Apart from the already mentioned 4C~12.50 (Mirabel
1989), detections of \HI\ absorption have been reported for 3C~433 (Mirabel
1989) and 3C~321 (Mirabel 1990).  These all happen to be among the most
powerful far-IR sources (for their range of radio power) and they all show
strong signs of starburst activity in the optical.  For example, 3C~433 is the
object which shows the clearest evidence for optical/UV starburst activity and
is also the most powerful object at far-IR wavelengths
in a survey of a complete
sample of 9 low redshift ($0.04 < z < 0.2$) 3CR radio galaxies (Wills et al.
in prep.).  Thus, although we do not have unbiased statistics, there seems to
be an effect related to the presence of a rich ISM that makes the detection of
\HI\ in absorption more likely.

As already pointed out by Mirabel (1982), there could be a link between
radio-IR spirals (like the starburst Arp~220) and radio galaxies in the early
stage of their evolution, like 4C~12.50.  In the formation of radio galaxies
and QSRs, the ultraluminous infrared galaxies (ULIRG) like Arp~220 are
believed to play an important role (see Sanders and Mirabel 1996 for a
review).  They may represent the early stage in a sequence of merging spiral
galaxies evolving through a starburst phase to eventually becoming QSOs and
powerful radio galaxies.  Thus, the starburst, far-IR galaxies that we are
considering could represent the result of such a process.  They should be
either young radio galaxies, as suggested by the fact that many of them are compact
and have a steep radio spectrum, or galaxies with re-started activity (like
3C~321 which has a steep spectrum core or obvious mergers as 3C~433).  A
number of ULIRGs have been detected in \HI\ absorption.  Their typical \HI\
column density is $>10^{21 - 22}$ cm$^{-2}$ for \tspin\ = 100 K.  Their broad
\HI\ profiles indicate rotation plus large amounts of unusually turbulent
neutral hydrogen (Mirabel 1982, Sanders \& Mirabel 1996).

\section{Conclusions}

We have presented \HI\ observations for 23 radio galaxies selected from 
the 2-Jy sample. \HI\ absorption has been detected in 5 objects.
We have investigated how the detection rate is distributed among galaxies with
different optical and radio properties.

For FR-I galaxies our data support the idea that
most of the \HI\ absorption in
these objects comes from a {\sl thin} nuclear disk, similar to those found in
other galaxies of this type (NGC 4261 and Hydra A).  Thus, this is consistent
with the HST imaging observations (Chiaberge et al.  1999) which do not
require the presence of pc scale, geometrically thick torus in low-luminosity
radio galaxies. 

The non-detection of \HI\ absorption in BLRGs is, to first order, consistent
with the unified schemes.  Broad line radio galaxies are in fact supposed to
be galaxies seen pole-on and therefore obscuration from the torus should not
occur.  

The fact that the \HI\ absorption is detected in NLRGs and not in the BLRGs
seems to suggest that a thick disk is indeed a cause of absorption. However,
it does not seem to explain all the observed characteristics of the
absorption.  For example, the detection of \HI\ absorption in far-IR bright,
starburst galaxies seems to indicate that there is also an effect related to
the presence of a rich ISM that makes detection of \HI\ in absorption more
likely.  We cannot say if the \HI\ is in a disk, but certainly in some objects
the situation seems to be more complicated than that.  In particular,
PKS~1814--63 and 3C~459 seem to have blueshifted absorption, possibly related
to some outflow.  PKS~1549--79 is likely to be a very young radio source and the
material around the nucleus has not yet been swept out by the expanding radio
source and/or quasar-induced winds.

It is clear that both high resolution VLBI data as well as accurate redshifts
are required for a more complete picture of what is producing the \HI\
absorption.

\section{Acknowledgements}

We wish to acknowledge Rene Vermeulen and Ylva Pihlstr\"om for comunicating
their results before publication.

\end{document}